\documentstyle[aps,epsfig,preprint]{revtex}
\draft

\author{Zafar Ahmed \\
Nuclear Physics Division, Bhabha Atomic Research Centre \\
Trombay, Bombay 400 085, India \\
zahmed@apsara.barc.ernet.in}
\title
{Handedness of complex PT-Symmetric potential barriers}
\date{\today}
\begin{document}
\maketitle
\begin{abstract}
Generally, when imaginary part of an optical potential is non-symmetric the
reflectivity, $R(E)$, shows left/right handedness, further if it is not
negative-definite the reflection and transmission, $T(E)$, coefficients become
anomalous in some energy intervals and absorption is indefinite ($\pm$).
We find that the complex PT-symmetric potentials could be exceptional in this
regard. They  may act effectively like an absorptive potential for any incident
energy provided the particle enters from the preferred (absorptive) side.
\end{abstract}
\vskip .2 in
When a potential is real the probability of quantal transmission, $T(E)$,
and reflection, $R(E)$, are invariant with respect to the side of incidence
of the particle. Recently, it has been proved [1] that when the potential is
complex and non-symmetric (in space), the value of reflectivity depends on
whether the particle is incident on the potential from left or right.
Thus when both the symmetries : time-reversal and parity are broken the
reflectivity shows handedness. For the scattering from a complex potential,
we have the condition of reciprocity satisfied
\begin{equation}
1- R_l(E)-A_l(E)=T(E)=1-R_r(E)-A_r(E)
\end{equation}
and unitarity is replaced by pseudo-unitarity
\begin{equation}
R(E)+T(E)+A(E)=1.
\end{equation}
Here $A$ denotes the probability of absorption and subscripts, $l$ and $r$,
stand for left and right. Let the scattering co-efficient, $S(E)$, represent
any of the probabilities : $R(E),T(E),A(E)$.
\par For a physical process the imaginary part of the optical potential
should be finite everywhere (it encloses a finite area).
Phenomenologically, it is the absorption (loss of flux into unknown channels)
which we try to study and model in a scattering process. We therefore require
$0<A(E)<1$ and this in turn requires both $R(E)$ and $T(E)$ to be non-anomalous
$(<1)$. Interestingly, this is achieved by choosing the imaginary part of the
potential as negative-definite in the whole regime. Notice that for an optical
potential $V_c(x)=V_r(x)+iV_i(x)$ that $A(E)$ is defined as
\begin{equation}
A(E)=-{2m \over \hbar^2 k}\int_{-\infty}^{\infty} V_i(x) \Psi^\ast(x) \Psi(x)~dx,
~V_i(x)\le 0,~V_i(\pm \infty)=0.
\end{equation}
A deeper study of Schr{\"o}dinger transmission from a complex potential reveals that
when the imaginary part is not negative-definite (e.g, it oscillates or it is
positive-definite), at least one of probabilities ($T(E)$, $R(E)$) in certain
intervals of
energy becomes anomalous ($>1$) and gives rise to indefinite ($\pm$) absorption
as a function of energy. Thus, such imaginary potentials have not been considered
and reported in scattering from one dimensional complex potentials.
\par Recently, a new scope for complex PT-symmetric $(P: x\rightarrow -x, T :
i\rightarrow -i)$ potentials to have real discrete eigenvalues [2-4] has been well
researched. Apart from real discrete spectrum, from such potentials real
energy-momentum dispersion relation have also been obtained [5-9] to find that
both usual and unusual energy bandstructure can exist/co-exist [8].
The question as to
what happens when particles are scattered from complex PT-symmetric potential
gains importance now. In the well explored area of barrier penetration
from one-dimensional
barrier potentials [11], one would like to know what happens when the potential
breaks time-reversal and parity symmetries individually but preserves them
jointly.
\par PT-symmetric potentials are expressible as $V(x)=V_e(x)+iV_o(x)$,
subscript
$e$ stands for $even$ and $o$ stands for $odd$. The imaginary part of
$V(x)$ being essentially, antisymmetric, consequently the (left/right)-handedness of
reflectivity [1] and anomalous values of $R(E)$ and $T(E)$ or equivalently
the indefiniteness of $A(E)$ as a function of energy are expected. Therefore,
these features also observed recently [10] do not actually subscribe only to
PT-symmetry or pseudo-Hermiticity of the optical potential. Complex PT-symmetric
potentials are found to be pseudo-Hermitian : $\eta H \eta^{-1}=H^\dagger$ [12].
\par In this Letter, we wish to report that
if a particle enters from the absorptive side of the complex PT-symmetric barrier,
we can still have non-anomalous and absorptive scattering at any energy :
$0 \le R(E),T(E )\le 1$ and $0 \le A(E) \le 1$.
This we refer to as handedness of a complex PT-symmetric potential.
\par Several interesting features of scattering from a one-dimensional potential
can be easily understood by constructing a novel rectangular potential as
\begin{equation}
V(|x|<a)=0, V(-a< x < 0) =V_0+is_1 V_2, V(0 < x < a)=V_0+is_2 V_2.
\end{equation}
and extracting $T(E)$ and $R(E)$ analytically. When $s_1=-1=s_2$, it will be
simple absorptive rectangular well. When $s_1=-1$ and $s_2=1$ it will be
PT-symmetric with absorptive side as on the `left'. When $s_1=1$ and $s_2=-1$
again
be PT-symmetric with absorptive side on the `right'. Non-PT-symmetric complex
potentials
with indefinite imaginary part can be had when $s_1 \ne s_2$ and $s_1 s_2 <0$.
Assuming the incidence of the particle from `left' we derive the scattering
co-efficients, $R^{p,q}_l(E)$, as
\begin{equation}
\left|{q(k^2-p^2)\sin pa \cos qa +p (k^2-q^2) \cos pa \sin qa+ik(p^2-q^2)\sin pa
\sin qa \over 2ikpq \cos pa \cos qa+p(k^2+q^2)\cos pa \sin qa+ q(p^2+k^2)\sin pa
\cos qa-ik(p^2+q^2) \sin pa \sin qa}\right|^2.
\end{equation}
Here $k=\sqrt{2mE}/\hbar,p=\sqrt{2m(V_1+is_1 V_2)}/\hbar,
q=\sqrt{2m(V_1+is_2 V_2)}/\hbar$.
Notice that $R^{p,q}_l(E)$ is not symmetric in $p$ and $q$ (when $s_1\ne s_2$)
displaying its left/right handedness [1]. The expression for $R^{p,q}_r(E)$ will
be nothing but
$R^{q,p}_l(E)$. The transmission co-efficient, $T(E)$ is given as
\begin{equation}
\left |{2ikpq
\over 2ikpq\cos pa \cos qa+p(k^2+q^2)\cos pa \sin qa+ q(p^2+k^2) \sin pa
\cos qa-ik(p^2+q^2) \sin pa \sin qa} \right |^2
\end{equation}
which is symmetric under the exchange of $p$ to $q$ displaying the
{\it reciprocity}
: its independence on the side of the incidence of particle. This invariance is
absolute whether or not $s_1=s_2.$ Let us fix $s_1=-1$ and $s_2=1$ so that the
imaginary part is absorptive on the left and next by computing from (5), very
remarkably, we find that $R_l(E) < R_r(E)$ and that $R_r(E)$ is anomalous but
$R_l(E)$ turns out to be physical $(<1)$.
\par Rectangular complex potential is more general than the complex PT-symmetric
potentials constructed from Dirac-delta potentials [6,8,10]. Interestingly, this
rectangular
potential (4) like its real counter-parts (rectangular well/barrier) will have
$S(E)$ as oscillatory. An interesting study reveals [13] that the rectangular
potential
is the most localized potential one can have and that other most commonly known
one-dimensional potentials  (e.g, Gaussian, Eckart, Lorentzian)
entail $S(E)$ as smooth function of energy. Consequent to this peculiar feature
of a rectangular potential, here we do not find parameters $V_1,V_2,a$
so that potential is absorptive for any energy excepting the situations
where $a$ is very small and the particle enters from the absorptive side.
\par We now take up PT-symmetric Scarf potential :
\begin{equation}
V(x)=V_1 \mbox {sech}^2(x/a)+iV_2 \mbox {sech}(x/a) \tanh (x/a).
\end{equation}
This we do, in order to demonstrate in a tractable setting that for a certain
choice of parameters,
a complex PT-symmetric potential would act as absorptive for any energy provided
the particle enters the potential
from the absorptive (preferred) side. Notice that  this potential is
absorptive (imaginary part is negative-definite) on the left ($x\rightarrow
-\infty$) and on the right hand, it is (imaginary part is positive-definite)
emissive. This potential is the first fully analytically solvable model of a
complex PT-symmetric potential entailing both discrete spectrum of both the
types : real when $V_2 \le {V_1+\Delta/4}$ and complex-conjugate pairs otherwise.
Thankfully, the complex reflection/transmission amplitudes for the real(non-complex)
Scarf potential have already been obtained by Khare and Sukhatme [14] in terms of
complex Gamma functions assuming the incidence from the left hand side.
Since the potential vanishes at $x=\pm \infty$ and is finite everywhere,
these can be extended to complex domain of the parameters. Very interestingly,
we find that Gamma functions give way to simple trigonometric (circular and
hyperbolic) functions in the expressions of $R(E)$ and $T(E)$. Thus, for the
complex PT-symmetric potential (7), we find
\begin{mathletters}
\begin{equation}
T(E)={2\sinh^2 2 \pi \kappa \over 2\cosh^ 2 2 \pi \kappa+ 4\cosh 2\pi \kappa
\cosh \pi f \cosh \pi g +\cosh 2\pi f+\cosh 2\pi g},
\end{equation}
\begin{equation}
F_l(E)={1 \over \sinh 2\pi \kappa}[e^{-\pi \kappa} \cosh{\pi f}+e^{\pi \kappa}
\cosh{\pi g}],
\end{equation}
\begin{equation}
R_l(E)= |F_l(E)|^2 T(E),
\end{equation}
\begin{equation}
A_l(E)=1-R_l(E)-T_l(E).
\end{equation}
\end{mathletters}
The parameters $\kappa$, $f$ and $g$ are defined as $\kappa =\sqrt{E \over \Delta}
=k a$, $f=\sqrt{{V_1+V_2 \over \Delta}-{1\over 4}}$, $g=\sqrt{{V_1-V_2 \over
\Delta}-{1\over 4}}$, where $\Delta=\sqrt {\hbar^2 \over 2ma^2}.$ When $V_1=V_2$, Eqs.
(8a) and (8b) become quite simple.
\par In this tractable setting, one can readily check that $T(E)$ (8a) is invariant
to the side of incidence of the particle by noticing that $T_l(E,V_2)=
T_l(E,-V_2)=T_r(E,V_2)$. Equivalently, $T(E)$ is invariant if $f,g$ are inter-changed.
Further, check that the transmission at any energy is non-anomalous : $0<T(E)<1$
as long as we have $V_1>0$ and $V_1>V_2-\Delta/4$. Generally, the imaginary part is a
perturbation of smaller magnitude in many physical processes, we readily check
that $R_l(E,V_2 : f,g)$ (8b) is always physical by remaining less than unity. However,
$R_r (E,V_2 : f,g) = R_l(E,-V_2 : g,f)$ becomes anomalous for smaller values of energy.
We also have $R_l (E) < R_r(E)$ provided $\Im (V_c(x<0))<0$ (see Eq. (8c)).
These results are displayed in Fig. 1.
\par When $V_2 >V_1-\Delta/4$, the hyperbolic functions of $g$ become trigonometric
and $T(E)$ attains anomalous values ($>1$) at energies around the barrier height
 : $E \approx V_1$. In this situation, we again find that both $R_l(E) < R_r(E)$ and
 $R_r(E)$ is anomalous from lower (sub-barrier) energies upto energies slightly
above the barrier height (see Fig. 2).
\par Apart from being analytically tractable, the Scarf potential is special
as it is shape (form) invariant in a certain setting [12]. In order to separate
out the presently claimed features from any other possible specialty of Scarf
potential, we study some analytically intractable complex PT-symmetric potential
models as illustrative examples. We consider two potentials  :
\begin{equation}
V(x)={V_1+iV_2 z \over (1+z^2)^4}
\end{equation}
and
\begin{equation}
V(x)=(V_1+iV_2 z)e^{-|z|},~z=x/a.
\end{equation}
These potentials being analytically intractable could be more general
as they do not belong to a known symmetry group.
However, a common feature among the potentials (7), (9) and (10) is that these are much
less localized in comparison to the rectangular potential.
\par Assuming $2m=1=\hbar$ and also $a=1$, we have computed the scattering
co-efficients by integrating the Schr{\"o}dinger equation numerically on both
the sides of $x=0$ for the models (7), (9), and (10). It may be important to
mention that the results (8a) and (8c) have also been checked against our
numerical integration method for the Scarf potential (7).
Various values of $V_1$ and $V_2$ (see Figs. 1-4) taken in arbitrary units are in
no way special excepting that we plan to address to two situations :
$V_2<V_2^{critical}$ and $V_2>V_2^{critical}$.
We find that there exists a characteristic critical
value of $V_2$ : $V_2^{critical}=f(V_1)$ above which even the transmission
probability becomes unphysical at energies around the barrier height,$V_1$,
(see Fig. 2) for a given complex PT-symmetric potential. For instance, for the
tractable Scarf potential (7), we have $V_2^{critical}=V_1-\Delta/4.$
The figures 1, 3 and 4 pictorially display the claimed feature of physical scattering
at least from one (absorptive) side of a complex PT-symmetric potential. For a given
complex PT-symmetric potential, we find that $V_2^{critical}$ may also be much higher
or  much lower than $V_1$. See Figs. 3 and 4 for the  potential models (9) and (10)
respectively, when $V_1=5.0$ and $V_2=4.0$, the scattering from left is physical.
We also find that when $V_1=4$ and $V_2=7.5$ the scattering from left remains
physical for the models (9) and (10). For these models, when $V_2=8.0$, $T(E)$ attains
anomalous values at energies around the barrier height, $V_1$, (as in Fig. 2).
The absorption in this case becomes indefinite ($\pm$) in certain intervals of energies
for the entrance of the particle from any side.
\par Complex PT-symmetric potentials surprise one readily as being non-Hermitian
and yet possessing a real discrete spectrum. Hence, equally dramatical speculation
that scattering from such potentials may yield no-absorption would not have been
very surprising. We find that this situation of no absorption does not
arise in the penetration through complex PT-symmetric potential barriers.
\par The other essence of a PT-symmetric potential governed by its parameters lies in
two kinds of results it yields : usual and unusual. For instance PT-symmetric
potentials give rise to real spectrum
with two branches [4], real spectrum with complex-conjugate pairs of eigenvalues
and usual [7,8] bandstructure containing discontinuous band gaps with unusual
[5-9] bandstructure also containing rounded bands. In this regard, the present work
reveals the handedness of a PT-symmetric optical potential barrier wherein the
scattering/tunneling will be physical ({\it usual}) only if the particle is incident
from absorptive (sink) side of the potential. If the particle enters from the other
side, the scattering will be anomalous ({\it unusual}) in certain intervals of energy.
Let us conclude by putting this feature in an amusing way : the complex PT-symmetric
potential barriers may mimic a `spy-glass' fitted in the windows of a room to view
out-side without being viewed from there. Model independent explanation of the
presently reported features may be very interesting.
\section*{References }
\begin{enumerate}
\item Z. Ahmed, Phys. Rev. A {\bf 64} (2001) 042716.
\item C.M. Bender and S. Boettcher, Phys. Rev. Lett. {\bf 80} (1998) 5243.
\item A. Khare and B.P. Mandal, Phys. Lett. A {\bf 272} (2000) 53;\\
G. Levai and M. Znojil, J. Phys. A : Math. \& Gen. {\bf 33} (2000) 7165;\\
B. Bagchi, S. Mallik, C. Quesne, R. Roychoudhury, Phys. Lett. A {\bf 289} (2001) 34;\\
G. Levai, A. Sinha, P. Roy, J. Phys. A : Math. \& Gen. 36 (2003)7611.
\item Z. Ahmed, Phys. Lett. A : {\bf 282} (2001) 343; {\bf 287} (2001) 295.
\item C.M. Bender, G.V. Dunne, P.N. Meissinger, Phys. Lett. A {\bf 252} (1999) 272.
\item H.F. Jones, Phys. Lett. A {\bf 262} (1999) 242.
\item J.K. Boyd, J. Math. Phys. {\bf 42} (2001) 15.
\item Z. Ahmed, Phys. Lett. A : {\bf 286} (2001) 231.
\item J.M. Cervero, Phys. Lett. A {\bf 317} (2003) 26.
\item R.N. Deb, A. Khare, B.D. Roy, Phys. Lett. A {\bf 301} (2003) 215.
\item J. Heading, J. Phys. A : Math. \& Gen. {\bf 6} (1973) 958;\\
P. Molinas-Mata and P. Molinas-Mata, Phys. Rev. A {\bf 54} (1996) 2060;\\
B. Sahu, I. Jamir, E,F. Lyngdoh and C. S. Sastry, Phys. Rev. C {\bf 57} (1998) 722;\\
J. Polao, J. G. Muga and R. Sala, Phys. Rev. Lett. {\bf 80} (1998) 5489;\\
M.V. Berry and D.H.J. O'Dell, J. Phys. A {\bf 31} (1998) 2093.
\item
 E.C.G. Sudarshan, Phys. Rev. {\bf 123} (1961) 2183;\\
 T.D. Lee and G.C. Wick, Nucl. Phys. B {\bf 9} (1969) 209;\\
 A. Mostafazadeh, J. Math. Phys. {\bf 43} (2002) 3944;\\
 Z. Ahmed, Phys. Lett. A {\bf 290} (2001) 19 ; A {\bf 294} (2002) 287.
\item Z. Ahmed (unpublished).
\item A. Khare and  U.P. Sukhatme, J. Phys. A : Gen. \& Math. {\bf 21} (1988) L 501.
\end{enumerate}
\begin{figure}[]
\psfig{figure=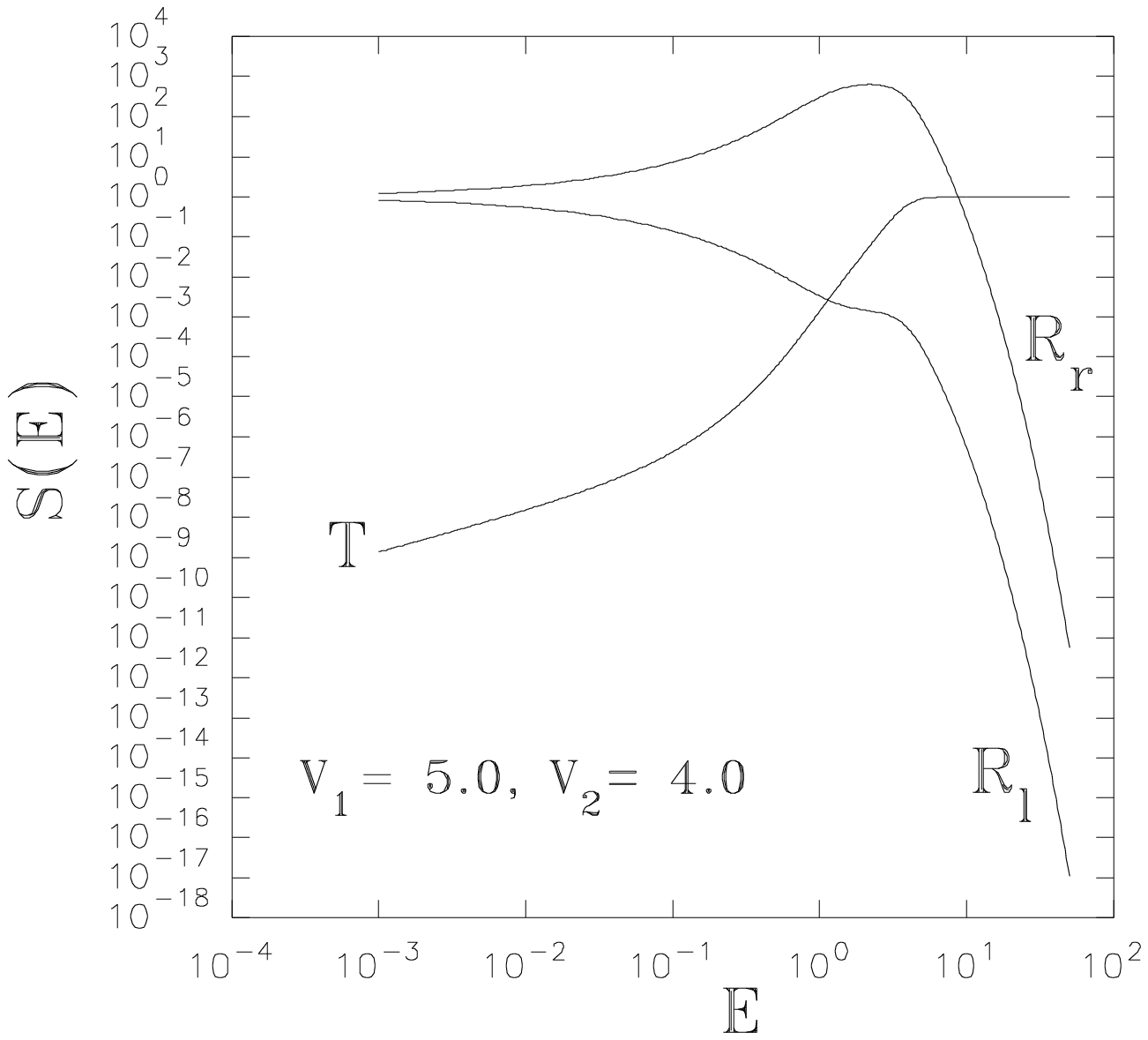,width=15.cm,height=9.cm}
\caption{Scattering co-efficients for the PT-symmetric complex Scarf potential
(7), notice that
$R_r(E)>R_l(E)<1$ and $0<T(E)<1$. Hence the scattering from left of the potential
is physical, i.e., $0<A_l(E)<1$.}
\end{figure}
\begin{figure}[]
\psfig{figure=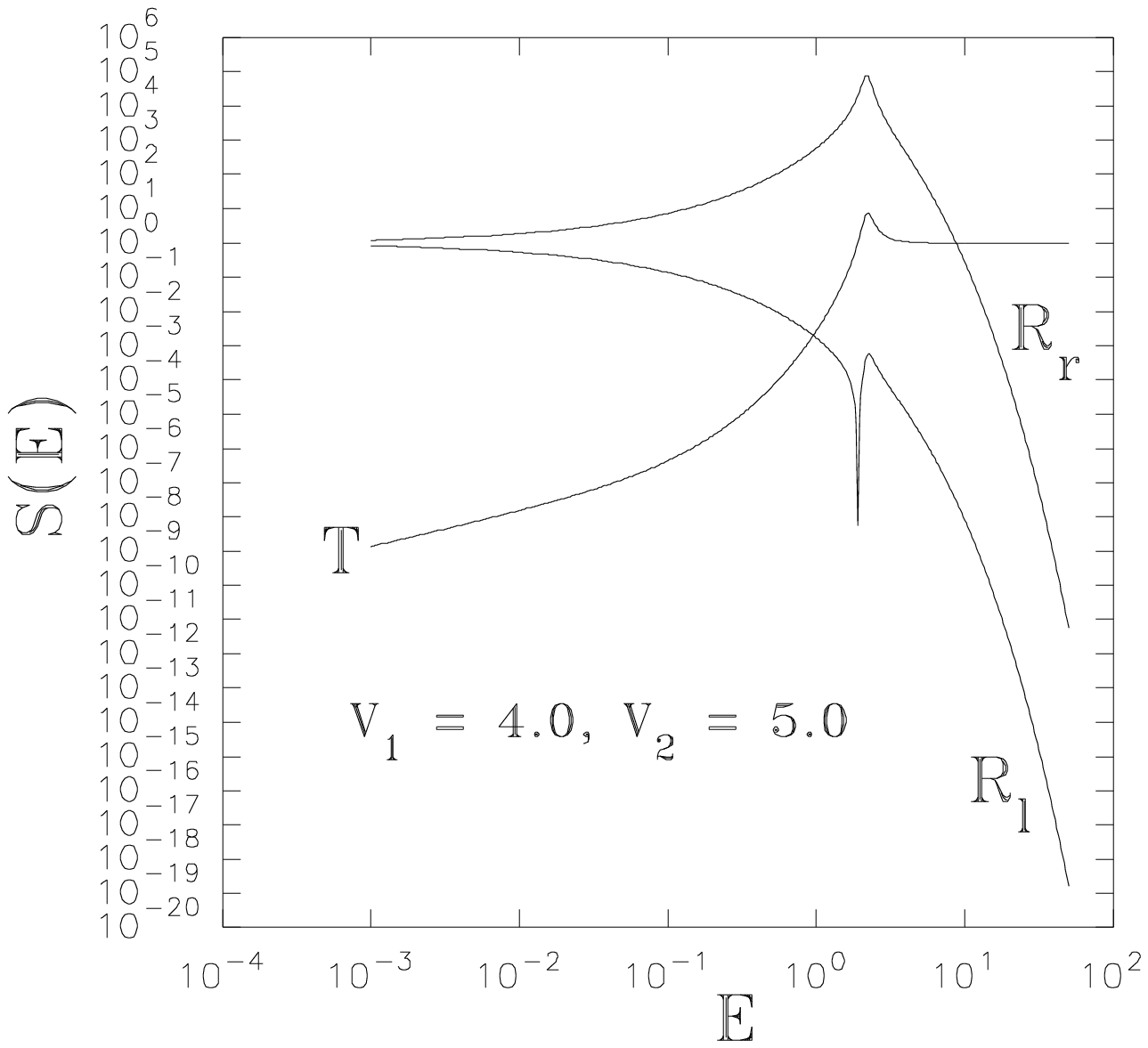,width=15.cm,height=9.cm}
\caption{When $V_2>V_2^{critical}=3.75$ in (7), $T(E)$ becomes
anomalous around the barrier height. Thus, scattering from either side will
yield an indefinite $(\pm)$ absorption as a function of energy.}
\end{figure}
\begin{figure}[h]
\psfig{figure=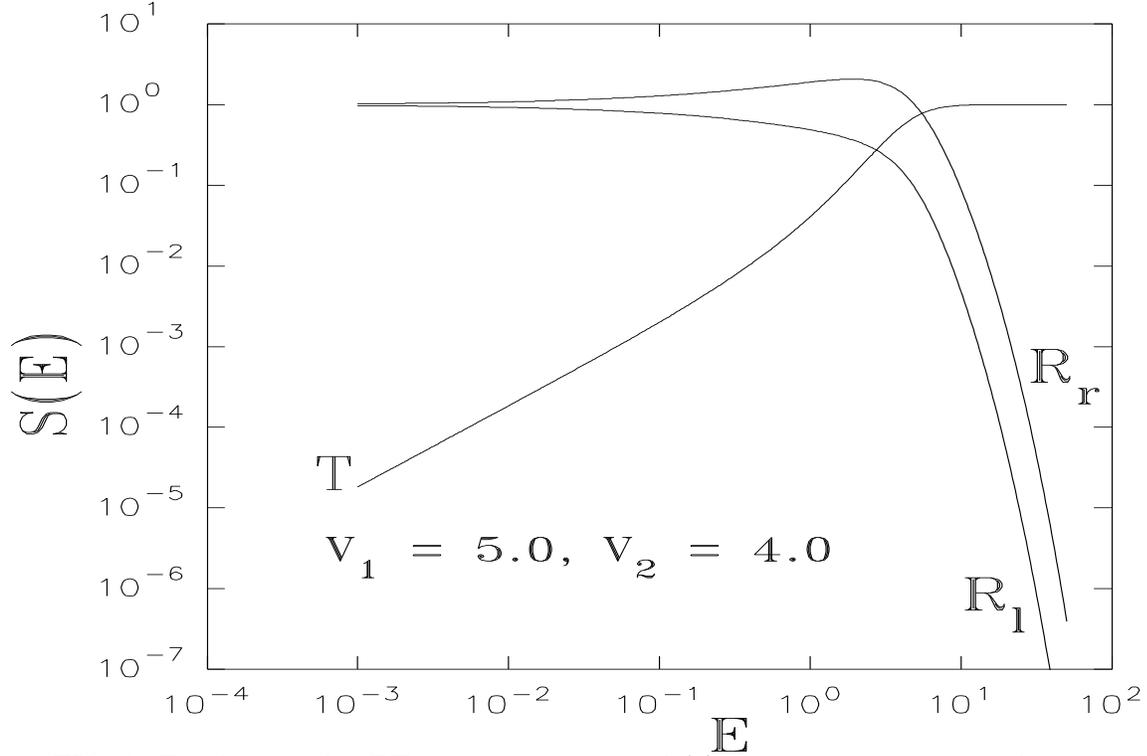,width=15.cm,height=10.cm}
\caption{For the complex PT-symmetric potential (9), for given parameters,
the scattering from left is physical. Notice that $0<T(E)<1$ and $R_r(E)>
R_l(E)<1$.}
\end{figure}
\begin{figure}[h]
\psfig{figure=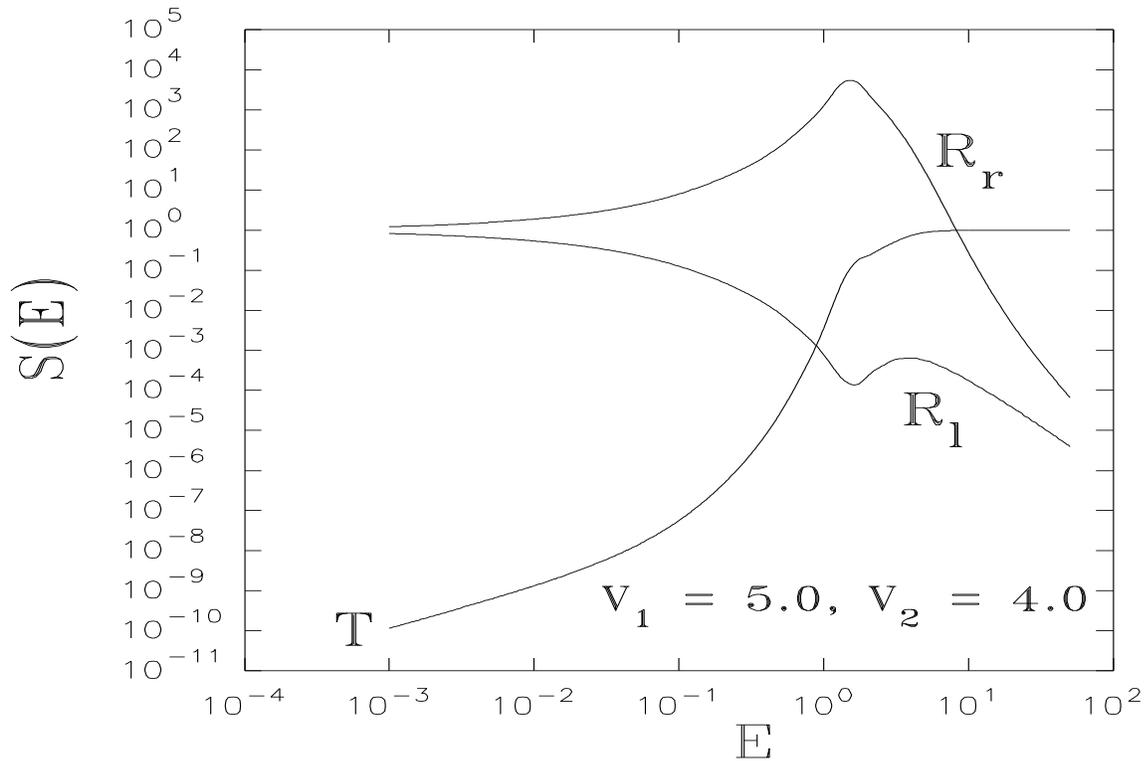,width=15.cm,height=10.cm}
\caption{The same as for Fig. 3 excepting that the potential is given by (10).}
\end{figure}
\end{document}